# Étude du comportement mécanique de préformes piquées

## *Evaluation of the mechanical behaviour of tufted preforms*

Imen Gnaba[1*], Peng Wang[1], Xavier Legrand[1] et Damien Soulat[1]

1 : Université Lille Nord de France, Gemtex
Ensait - École Nationale Supérieure des Arts et Industries Textiles
2 Allée Louise et Victor Champier, BP 30329-59056 Roubaix Cedex 1 FRANCE
*e-mail : imen.gnaba@ensait.fr

**Résumé**

Les faibles propriétés dans l'épaisseur des stratifiés en termes de rigidité, de résistance à la fatigue et à l'impact, impliquent dans le cadre des pièces épaisses une sensibilité au délaminage. Cette problématique a conduit à l'émergence de préformes tridimensionnelles (3D) disposant de renfort dans l'épaisseur permettant ainsi de combiner de bonnes performances mécaniques dans le plan et hors plan. Dans cette étude, la technologie de renforcement par piquage dans l'épaisseur est utilisée. Cette technique de renforcement est utilisée dans le domaine des matériaux composites, en raison de leur potentiel en termes de renforcement de préformes épaisses. Dans le cadre de cette étude, on se focalise sur le comportement des préformes renforcées par piquage, avant imprégnation, et ce pour leurs utilisations lors des étapes de mises en forme lors des procédés de fabrication. Pour ce faire, des essais élémentaires de caractérisation mécanique en traction, flexion et cisaillement plan, ont été menés. Les résultats mettent en évidence l'influence du piquage sur les propriétés mécaniques du renfort.

**Abstract**

The weak properties in the thickness of the laminates in terms of rigidity, fatigue and impact resistance involve delamination sensitivity for thick structures. This issue has led to the development of three-dimensional (3D) preforms with reinforcement through-the- thickness, making it possible to combine good mechanical performance in-plane and out-of-plane. In this study, the tufting technology is used to strength through-the-thickness of stacked layers. This reinforcement technique has extensively gained attention in the composite industry due to its ability to reinforce thick pieces. The present study focuses on the mechanical behaviour of tufted preforms before resin impregnation due to their application in the forming steps during the manufacturing process. Therefore, an experimental investigation of the mechanical behaviour (tensile, bending and in-plane shear) was carried out in order to outline the influence of tufting on the mechanical properties of the final structure.

**Mots Clés :** piquage, renforcement dans l'épaisseur, traction, flexion, cisaillement plan

## 1. Introduction

La technologie de piquage été auparavant utilisée uniquement pour la fabrication de moquettes, tapis et vêtements thermiques. Cette technologie fût actuellement un axe de développement majeur pour renforcer les structures épaisses à géométrie complexe [1].

Le piquage dérive du principe de la couture classique comme l'illustre la figure 1 (fig. 1). Cependant, ce procédé ne requiert l'accès qu'à une seule de la préforme et peut-être réalisé à l'aide que d'seul fil de liage continu. Ces avantages facilitent la production de structures piquées et favorisent l'automatisation du procédé de piquage [2-4].

En effet, le fil de piquage est introduit par l'intermédiaire d'une aiguille creuse sans générer de tension à la surface du stratifié, ce qui permet d'éviter le cisaillement ainsi que l'ondulation des fils au sein de la structure. Les boucles sont alors créées avec un simple frottement du fil, qui est maintenu à l'intérieur de la préforme, lors du retour de l'aiguille.

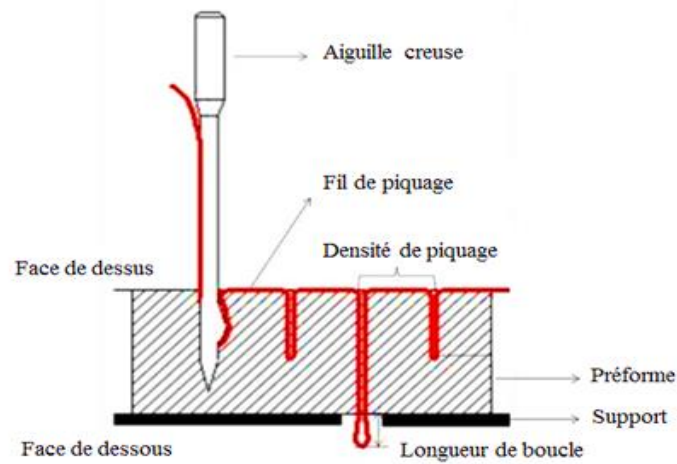

*Fig. 1. Principe de piquage.*

## 2. Matériaux

Afin de produire des préformes piquées, un banc de renforcement par piquage (fig. 2) a été élaboré et automatisé au sein du laboratoire GEMTEX.

La présente machine est équipée par les éléments suivants :

- Une tête de piquage pouvant se déplacer dans les deux directions du plan, et sur laquelle une aiguille, de 2 mm de diamètre, est positionnée. L'aiguille est liée à des vérins pneumatiques permettant le contrôle de la profondeur de piquage. Le choix de l'aiguille dépend de la nature de la préforme ainsi que des caractéristiques du fil de piquage.

- Un pied presseur est actionné par le biais d'un autre vérin pneumatique permettant à la fois de maintenir la préforme et d'exercer une certaine pression lors du procédé de piquage.

- Un dispositif d'alimentation (bobine) permettant de fournir le fil de piquage avec une certaine tension et longueur.

- Un bâti mobile permettant le mouvement simultané des différents équipements.

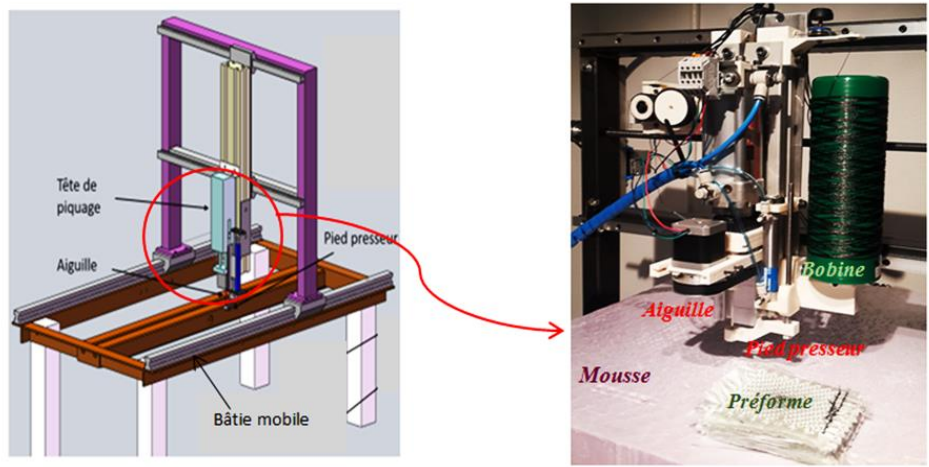

*Fig. 2. Dispositif de piquage développé au GEMTEX.*

L'étude expérimentale, présentée dans ce travail, a été réalisée à l'aide d'un tissu de carbone commercial destiné à des applications aéronautiques, de référence « HexForce 46290 WB 1010 ». Le tissu est constitué de fibres de carbone « HexTow IM7 GP 6K » en chaîne et en trame.

L'armure du tissu est un satin de 5 de contexture nominale de 6,5 fils/cm en chaîne et 6,5 coups/cm en trame avec une masse surfacique de 290 g/m² et une épaisseur de l'ordre de 0,3 mm.

En ce qui concerne le fil de piquage, il s'agit d'un fil de carbone réalisé à partir de la fibre Tenax JHTA 40 E15/ 2K 67 Tex 15 S produit par la société Schappe techniques. Il est à noter que le fil choisi doit être compatible avec la taille de l'aiguille utilisée.

Les caractéristiques des différentes structures renforcées par piquage sont détaillées dans le tableau 1 (tab. 1). Le piquage est caractérisé par un pas de piquage, un motif relativement aux directions chaîne et trame des couches, et une longueur de boucles. Dans le cadre de cette étude le piquage est perpendiculaire et traversant.

| Références | Renfort | Fil de piquage | Nombre de couches | Pas de piquage [mm] | Longueur des boucles [mm] | Motif de piquage |
|---|---|---|---|---|---|---|
| Non-piqué | Carbone (*Satin 5*) | - | 4 | - | - | - |
| Piqué 0° (sens trame) | | Carbone (*2 x 1K 67 Tex 15 S*) | | 10 | 25 | 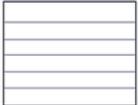 |
| Piqué 90° (sens chaîne) | | | | | | 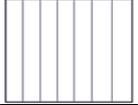 |
| Piqué 0°/90° (croisé) | | | | | | 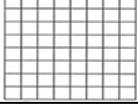 |

*Tab. 1. Caractéristiques des préformes testées.*

### 3.    Méthodes de caractérisation

#### 3.1  Caractérisation en traction

Le comportement en traction uni-axiale est caractérisé selon le protocole de la norme EN ISO 13934-1où l'éprouvette est placée entre les deux mors à une distance égale à 200 mm et sollicitée avec une vitesse de 100 mm/min. Des talons de 50 mm x 50 mm sont collés aux extrémités de l'éprouvette (hors zone utile) afin d'éviter toute sorte de déchirement et de glissement du renfort au niveau des mors. Les éprouvettes testées se caractérisent par une largeur de 50 mm et une longueur utile de 200 mm et l'essai s'effectue dans le sens chaîne. Afin d'avoir un comportement en traction plan assez répétitif, cinq éprouvettes par structures ont été testées.

#### 3.2  Caractérisation en flexion

L'essai de flexion permet d'évaluer la rigidité en flexion lors du tombé sous propre poids des renforts.

La rigidité en flexion est déterminée par le biais du dispositif KES (fig. 3). Il s'agit d'un simple plan incliné à 41,5° sur lequel le renfort textile est déposé puis est laissé glisser tout au long du plan. A l'aide d'une règle graduée, on mesure la longueur suspendue lorsque l'extrémité de l'éprouvette touche le plan incliné. Le comportement en flexion a été identifié selon le protocole de la norme

ISO 4604 où cinq mesures par structures ont été réalisées afin de déterminer le comportement typique en flexion.

Les rigidités de flexion (G, mN.m) des structures piquées et non-piquées ont été calculées à partir de la longueur moyenne de tombée de l'échantillon (L, m) ainsi que sa masse surfacique (ρ, g/m²) comme indiqué dans l'équation 1 (Eq. 1) [5].

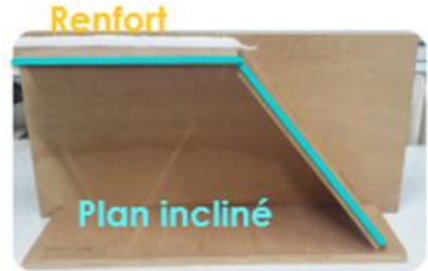

*Fig. 3. Dispositif « KES ».*

$$G = 9{,}81 \ \rho \ \left(\frac{L}{2}\right)^3 \qquad \text{(Eq. 1)}$$

### 3.3 Caractérisation en cisaillement

Le cisaillement plan est reconnu comme étant la déformation principale lors de l'étape de préformage des tissus avant de procéder à l'imprégnation de la résine, dans les procédés de fabrication des renforts de type LCM.

Le comportement en cisaillement plan est déterminé, dans cette étude, par un essai dit de « Bias-test » [6-8]. Dans cette expérience, un essai de traction est effectué sur une éprouvette rectangulaire dans laquelle les directions chaîne et trame des mèches sont orientées initialement à 45° par rapport au sens de sollicitation.

Les éprouvettes testées doivent présenter un rapport de longueur sur largeur supérieur à deux afin de disposer d'une zone de cisaillement, où les renforts ne sont pas pris entre les mors. Dans ces travaux, les éprouvettes ont une longueur utile de 210 mm pour une largeur de 70 mm et la vitesse de l'essai est de l'ordre de 30 mm/min.

Au cours d'un essai de bias-test, trois zones spécifiques peuvent être distinguées (fig. 4) :

- **Zone A** : zone non-déformée où les mèches sont considérées comme inextensibles et qu'aucun glissement ne se produit au niveau des extrémités de l'éprouvette.

- **Zone B** : zone demi-cisaillée où le réseau fibreux est encastré à une seule extrémité.

- **Zone C** : zone de cisaillement pur où les mèches demeurent libres à leurs extrémités. L'absence de glissement ainsi que l'élongation suffisante de l'éprouvette conduisent à une pure déformation en cisaillement qui se caractérise par un angle de cisaillement. Une analyse géométrique de l'essai [8], permet de calculer l'angle de cisaillement (γ, °) en fonction du déplacement des mors (d, mm) et la taille initiale de l'éprouvette (D, mm) comme le montre l'équation 2 (Eq. 2).

$$\gamma = \frac{\pi}{2} - 2\cos^{-1}\left(\frac{D+d}{\sqrt{2}\,D}\right) \qquad \text{(Eq. 2)}$$

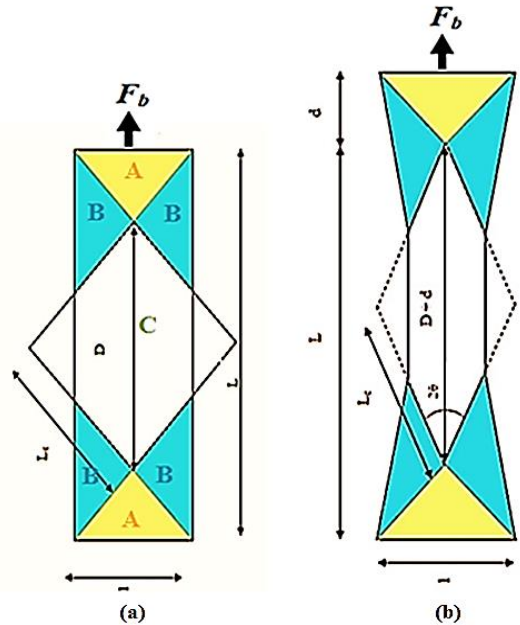

*Fig. 4. Allongement d'une éprouvette du bias-test (a) : état initial, (b) : état déformé [9].*

## 4. Résultats expérimentaux

### 4.1 Traction plan

Les courbes moyennes d'effort en fonction de déformation des préformes piquées et non-piquées sont présentées dans la figure 5 (fig. 5). D'un point de vue global, on remarque que toutes les courbes effort/déformation se caractérisent par une non-linéarité de comportement pour les faibles efforts.

A l'état sec, le piquage influe sur le comportement en traction plan où toutes les préformes piquées se caractérisent par une augmentation de la déformation à la rupture et une diminution de la force maximale, ce qui entraîne une rigidité plus faible du matériau.

Les courbes présentées peuvent être divisées en trois zones principales :

- **Zone d'alignement** : Dans cette première partie des courbes, l'effort appliqué aligne le réseau fibreux dans la direction de sollicitation en éliminant l'embuvage. Il est à noter que cette zone est plus importante pour les préformes piquées. Les fils de renforcement par piquage ayant tendance à freiner cette dés-ondulation.

- **Réponse linéaire** : Dans cette partie des courbes les renforts, sens chaîne, sont alignés dans la direction de sollicitation, la réponse est linéaire. On constate pour les préformes piquées que la pente dans cette zone est plus faible que la préforme non piquée.

- **Zone de rupture** : Dans cette zone, caractérisée par la chute de l'effort, le tissu de carbone est entièrement endommagé où la rupture est engendrée par la détérioration du réseau fibreux sans aucun endommagement de fil de piquage. Les boucles sont avalées uniquement au niveau de la zone de rupture comme illustré dans la figure 6.

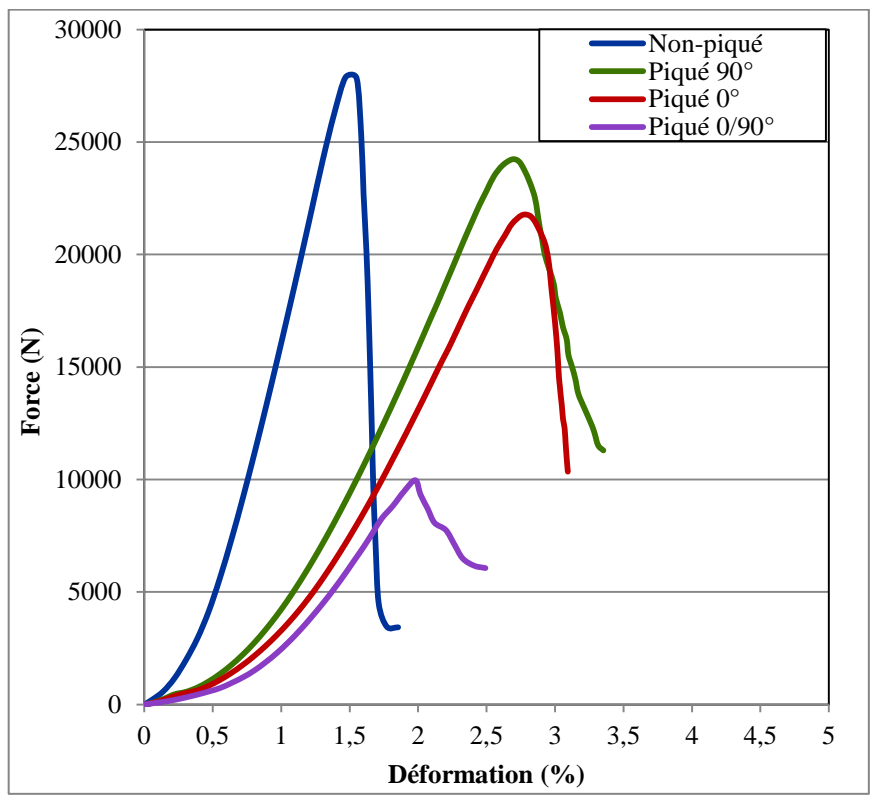

*Fig. 5. Courbes moyennes en traction plan des préformes testées.*

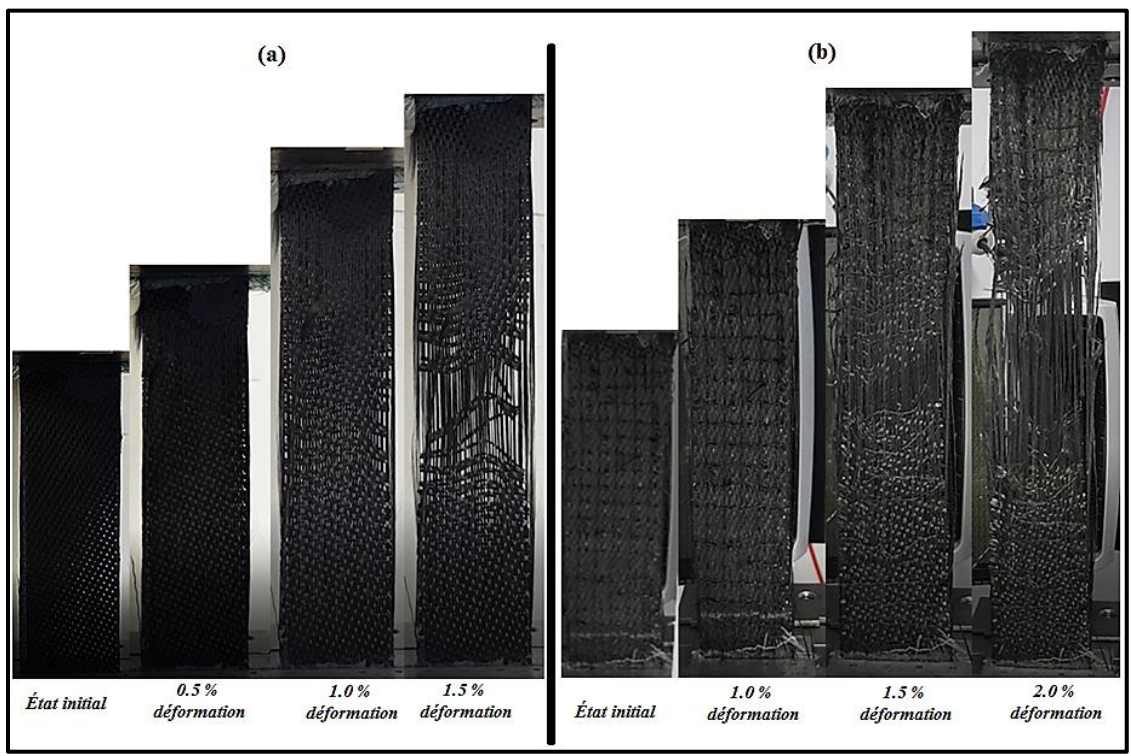

*Fig. 6. Exemple des éprouvettes en traction plan (a) : non-piqué, (b) : piqué 0°/90°.*

Des courbes de la figure 5, les caractéristiques en traction des préformes testées sont extraites et rapportées dans le tableau 2 (Tab. 2).

Les résultats démontrent que le renfort piqué dans le même sens que la charge appliquée (piqué 90°) présente une résistance plus élevée que les autres préformes piquées (à 0° et à 0°/90°).

La préforme piquée à 0°/90° présente des propriétés mécaniques les plus faibles suite au double motif de piquage (piquage dans le sens chaîne et trame) qui peut endommager le réseau fibreux.

Toutes les préformes piquées se caractérisent par une déformation à la rupture supérieure à celle du renfort non-piqué.

| Références | Force maximale [N] | Déformation à la rupture [%] |
|---|---|---|
| Non-piqué | 27883,7 | 1,47 |
| Piqué 0° | 21745,6 | 2,8 |
| Piqué 90° | 24233,7 | 2,7 |
| Piqué 0°/90° | 9967,9 | 2,0 |

*Tab. 2. Propriétés en traction plan des préformes testées.*

**4.2 Comportement en flexion**

L'illustration de la Figure 7 (fig. 7) présente les valeurs des rigidités de flexion des préformes piquées et non-piquées. Les résultats montrent que la présence d'un renforcement dans l'épaisseur influe significativement le comportement en flexion où toutes les structures piquées, indépendamment du motif de piquage, présentent une rigidité en flexion plus élevée que la préforme non-piquée.

Les travaux présentés révèlent que le piquage tend à améliorer la raideur en flexion, cependant des essais complémentaires sont nécessaires afin de valider cette affirmation.

La rigidité en flexion est principalement contrôlée par l'évolution de la masse surfacique où la rigidité est proportionnelle à la masse surfacique.

La préforme orientée à 0°/90° présente la rigidité en flexion la plus élevée de l'ordre de 29,6 mN.m. Ceci est expliqué par le double piquage, dans les sens chaîne et trame, où on rajoute plus de la matière (plus de fil de piquage) ce qui entraîne une augmentation de la masse surfacique.

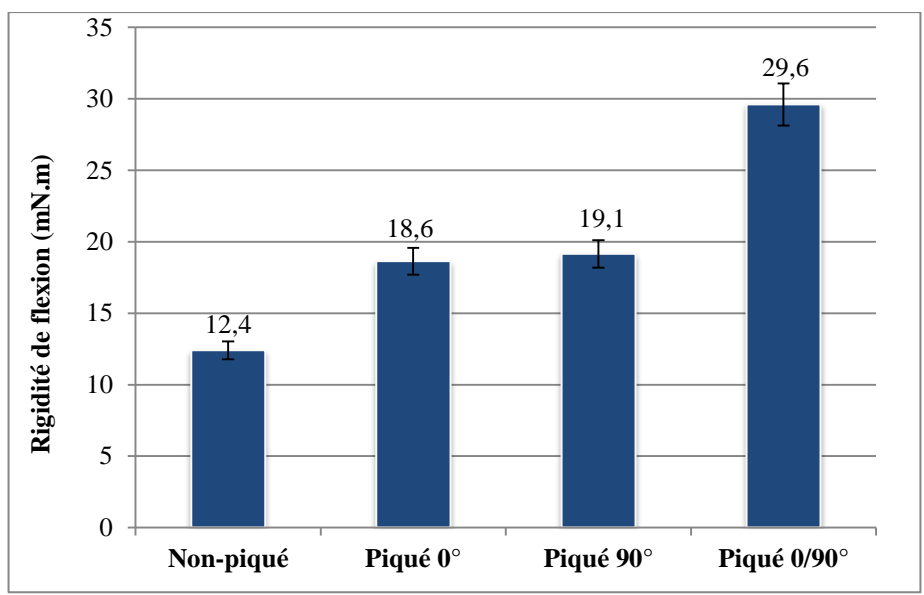

*Fig. 7. Raideur en flexion des structures piquées et non-piquées.*

**4.3 Comportement en cisaillement plan**

A partir des essais de bias-test, il est possible d'extraire les principales propriétés mécaniques d'un comportement en cisaillement plan en termes d'effort, de déplacement et d'angle de cisaillement.

La figure 8 (a) regroupe les courbes de la force en fonction de déplacement, issues de l'essai de cisaillement plan, des préformes piquées et non-piquées.

Les réponses des différentes éprouvettes possèdent une allure identique où la résistance du matériau augmente progressivement en fonction du déplacement et trois zones peuvent être distinguées :

- **Zone 1** : caractérisée par une faible rigidité du matériau qui se traduit par une augmentation circonstanciée des déformations à très faible effort. Cette zone est associée à la rotation des renforts fibreux.

- **Zone 2** : caractérisée par une rigidification progressive due à la rotation avec frottement du réseau fibreux. Dans cette région, les fibres sont fortement sollicitées dans la direction de l'effort appliqué.

- **Zone 3**: correspond à la rigidité en cisaillement pur où les contraintes au sein du renfort sont très élevées. Dans cette zone, le tissu de carbone est dit « bloqué ».

La préforme piquée dans la même direction de l'essai, cas du piqué 90°, montre un phénomène de tension au niveau des points de piquage qui conduit par la suite à une augmentation plus importante de la force de cisaillement. On constate sur ces courbes en effort/déplacement que les préformes piquées démarre plus tôt la rigidification en cisaillement comparativement à la préforme non-piquée

Les courbes moyennes de l'angle de cisaillement en fonction du déplacement sont présentées dans la figure 8 (b) où l'angle a été mesuré au niveau de la zone de cisaillement pur (zone C fig. 4). Le comportement obtenu est assez répétitif pour les différentes structures testées. La courbe théorique de l'évolution de l'angle de cisaillement en fonction du déplacement (Eq.2) est superposée à ces relevés expérimentaux.

A un angle donné, les courbes expérimentales des préformes piquées divergent de la courbe théorique. Ce point correspond à l'angle de blocage qui est aux alentours de 55°.

Cependant, la courbe de la préforme non-piquée se sépare plus vite de la courbe théorique par rapport aux renforts piqués.

En conséquence, il est constaté que la liaison créée par le piquage bloque le glissement entre les réseaux fibreux et par la suite la rotation du matériau.

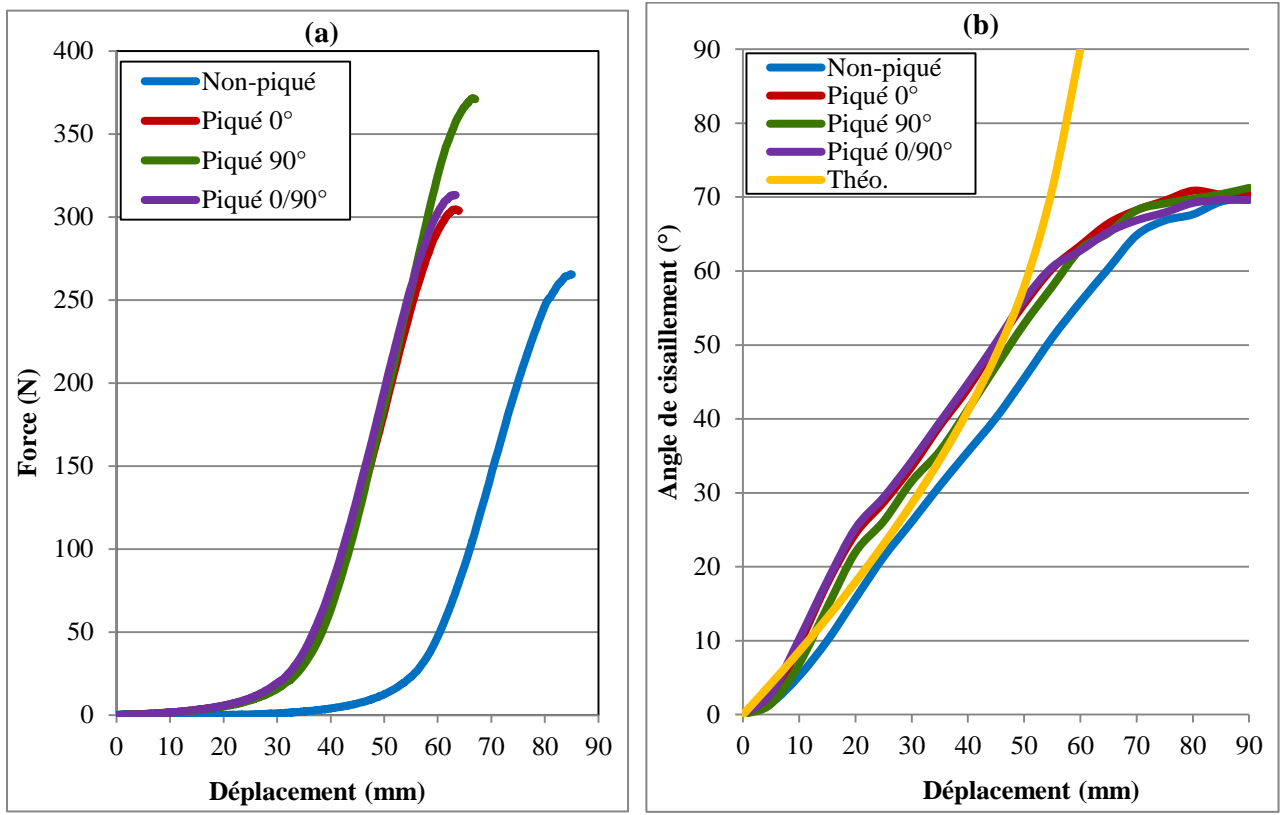

*Fig. 8. Courbes de cisaillement (a) : Force/Déplacement, (b) : Angle de cisaillement/Déplacement.*

De plus, des observations visuelles des préformes piquées et non-piquées ont été conduites lors d'un essai de bias-test afin de visualiser l'évolution des éprouvettes à différents déplacements (fig. 9) (d représente le déplacement au cours de l'essai).

Les évolutions des éprouvettes révèlent une déformation significative des échantillons piqués et non-piqués. Ces figures mettent en évidence les modes de déformation des différentes zones de l'éprouvette.

La déformation « du pied » des éprouvettes met en évidence un phénomène de glissement intra-plis relativement au cisaillement. Ce phénomène est localisé à la jonction des différentes zones de l'éprouvette tout en générant une courbure des mèches. Ce phénomène apparaît d'une manière symétrique sur les deux faces de l'éprouvette.

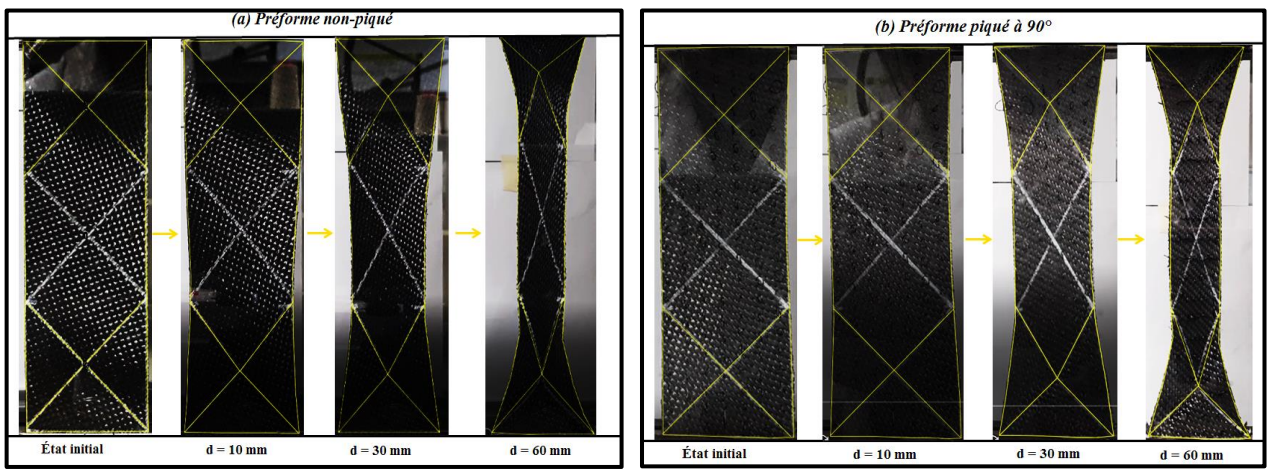

*Fig. 9. Evolution en bias-test (a) : éprouvette non-piquée, (b) : éprouvette piquée à 90°.*

## 5. Conclusion

La technologie de piquage est utilisée pour le renforcement dans l'épaisseur en raison de sa faculté à lier plusieurs couches ensemble en ayant accès qu'à une face de l'empilement. Cependant, cet ajout d'un renforcement dans l'épaisseur peut engendrer une dégradation des propriétés mécaniques dans le plan où une étude approfondie du processus de piquage ainsi que les paramètres impliqués est préconisée afin de comprendre leur influence. Dans la présente étude, une approche expérimentale sur l'influence de piquage sur le comportement mécanique dans le plan a été menée.

Le comportement en traction plan démontre que les structures piquées se caractérisent par une déformation à la rupture plus élevée où les fils de piquage diffèrent le phénomène de rupture. Par contre cette étude montre qu'en traction, l'effort à la rupture est plus faible pour les structures piquées où on peut conclure que le piquage tend à endommager le réseau fibreux. De plus, une orientation de piquage qui suit la direction de l'essai présente une résistance du matériau plus élevée.

Concernant le comportement en flexion, l'ajout des fils de renforcement affecte significativement la rigidité en flexion du matériau. Cette rigidité en flexion est principalement contrôlée par la masse surfacique où un rapport de proportionnalité existe entre ces deux paramètres. Le comportement en cisaillement plan souligne des propriétés mécaniques assez pertinentes des préformes piquées où une augmentation de l'effort de cisaillement a été enregistré pour la structure piquée dans le même sens de sollicitation. Il a été aussi démontré que la présence des points de piquage bloque, en termes d'angles, le glissement entre les réseaux fibreux et par la suite la rotation du matériau.

L'étude de la déformabilité des renforts piqués sur un banc de préformage, est l'objectif de nos prochains travaux où une analyse des différents paramètres (type de poinçon, pression serre-flan, empilements…) est prévue. Il est important de développer aussi la modélisation numérique et la simulation de l'emboutissage des structures piquées afin de prédire les conditions réalisables de préformage.

**Remerciements**

**Références**